\documentclass[pra,showpacs,uplatex]{revtex4}

\usepackage{amsmath}
\usepackage{bm}
\usepackage[dvipdfmx]{graphicx}
\usepackage[dvipdfmx]{color}
\usepackage[dvipdfmx]{epsfig}
\usepackage{float}
\usepackage{ulem}
\usepackage{color}

\begin{document}

\title{Evolution of the Stokes parameters, polarization singularities and optical skyrmion}

\author{Hiroshi Kuratsuji}
\email{kra@se.ritsumei.ac.jp}
\address{Office of professor emeritus, Ritsumeikan University-BKC, Kusatsu City, Shiga, 525-8577, Japan}

\author{Satoshi Tsuchida}
\address{Department of Physics, Osaka City University, Osaka, Osaka, 558--8585, Japan}

\date{\today}

\begin{abstract}
A physical theory is presented for polarized light from an aspect of polarization singularity.
This is carried out by analyzing  the  evolution equation of the Stokes parameters
that is derived from the nonlinear Schr\"odinger type equation.
The problem is explored from two aspects:
The first is concerning the single mode Stokes parameter that is described by the propagation distance $ z $.
The trajectory of the polarization singularities is simply given by the points satisfying
a rule: $ S_3 (z) = \pm S_0 $ and  $ 0 $
($ S_3 $ represents the third component of the Stokes parameters), which form a tube like surface
centered with the C-line surrounded by L-surfaces in the presence of the linear as well as the nonlinear birefringence.
It is pointed out that the optical activity (or chirality) plays an ``obstacle" to reveal the polarization singularity.
As the other aspect, we consider the field theoretic aspect of the Stokes parameters, for which a special solution,
called an optical skyrmion, is constructed to explicate the polarization singularity.
The skyrmion forms a {\it tube}, and the dynamical content of which is briefly discussed. On the basis of 
the work  presented  here further analysis  would be expected to reveal  hidden aspects  of polarization optics.
\end{abstract}

\maketitle

\section{Introduction}

The zero point of complex wave function plays a peculiar role in physics.
Among others, the striking consequence is an emergence of a nodal singularity accompanying
the quantized monopole~\cite{Dirac}.
Now what happens if this is extended  to a complex wave consisting of two components?
A typical problem is the light polarization~\cite{LL,BW},
which is described by two-component wave or spinor.
Hence one expects that a concept of the singularity is hidden in the light polarization.
Indeed the polarization singularity was recognized in Refs.~\cite{Nye1,Nye2}
and thoroughly discussed in the articles~\cite{Dennis,Berry}.
The essence is that; the singularities are given by
two kind of objects; one is the circular polarization point (or line) which is named the C-point or line,
and the other is the linear polarization line (or surface) called L-line or surface.
The C-point is given by the north (or south) pole on the Poincar\'e sphere representing the Stokes parameters.
As is well known, if the point on a sphere is coordinated by the angular parameters $ (\theta, \phi) $,
we have difficulty  defining  the angle $ \phi $ at both poles.
Namely, the axis of rotation loses its meaning at the poles, which corresponds to circular polarization.
As for the linear polarization, the situation is a little bit simple:
An equatorial circle divides the Poincar\'e sphere into two hemisphere and
by this division the chirality of polarization suddenly changes.

In the works of Refs.~\cite{Dennis,Berry} great detail of the behavior of polarization singularities
has  been given with special emphasis on the topological classification of the singularities.
More specifically, in Ref.~\cite{Dennis}, the polarization singularities were analyzed in paraxial field with
applications to morphology and statistics, and in Ref.~\cite{Berry},
the optical singularities of birefringent dichroic chiral crystals were analyzed. 
Furthermore, the structure of line of circular and linear 
polarization in three dimensional ellipse fields was given in Ref.~\cite{Freund}.

However, apart from such morphological analysis,
it is intriguing to address the problem  of how the singularity evolves during the propagation through anisotropic media.
To explore this topic is the aim of the present article.
This is carried out by analyzing the evolution equation of the Stokes parameters,
which is based on the Maxwell equation that is expressed in terms of the paraxial scheme
assisted by the envelope approximation.

We treat the problem from two aspects;
one is the single mode polarized wave and the other is the field theory aspect as is explained below.
In Sec.~\ref{sec:preliminary} and ~\ref{sec:trajectory},
we consider the single mode Schr\"odinger type equation,
which is described by  the single coordinate $ z $, i.e., the propagation distance, while the Hamiltonian depends
on the transversal coordinate $ ( x , y ) $ (see Sec.~\ref{sec:preliminary}).
This is transcribed to the equation of motion for the single Stokes parameters~\cite{Sala, Tratnik, Kakigi}.
Here the crucial point is that the coordinates $ (x, y) $ should be treated only 
as parameters which control the polarization singularity.
Subsequently, in Sec.~\ref{sec:trajectory} an analysis of the Stokes parameters is given
in the presence of the linear and nonlinear birefringence which are induced by the external
Kerr effect and/or nonlinear coupling of polarization components~\cite{Sala,Kra}.
The trajectory of  the C-point, which is just the C-line, is simply given by the point set satisfying $ S_3 (z) = \pm S_0 $.
Similarly, the L-surface, which is traced by the L-line, corresponds to the point set satisfying $ S_3 (z) = 0 $.
Here $ S_3 $ means the third component of the Stokes vector. It is important to mention here 
that the optical activity gives rise to a difficulty in controlling  the polarization singularity.

In Sec.~\ref{sec:skyrmion}, as a complementary aspect to the single mode,
we consider the field equation for the Stokes parameters,
which is arranged so as to include the distribution over the transversal plane $ ( x, y ) $~\cite{Snyder,Swartz,Rozas}.
Within this framework the polarization singularity is treated as a special solution,
similar to the ``skyrmion"~\cite{Skyrme}.
This object implies a common viewpoint that could be shared with the trajectory
of polarization  singularity for the single mode Stokes parameters.
A dynamical content of the optical skyrmion is briefly discussed.

\section{Preliminary}
\label{sec:preliminary}

\subsection{The Stokes parameters in paraxial scheme}

Here we adopt the well known procedure of the ``paraxial approximation''.
Let $ {\boldsymbol{E}} $ be the light field (electric field) which is traveling
through the anisotropic medium in $ z $-direction.
We suppose the modified plane wave~\cite{LL}:
$ {\boldsymbol{E}} (x, y, z) = {\boldsymbol{f}} (x, y, z) \exp[ i k n_{0} z ] $
where $ k = \frac{ {\omega} }{ c } $ and $ n_0 ( \equiv \sqrt{ \epsilon_{0} } ) $
is the refractive index of the isotropic medium that is just used for reference.
We choose the $ z $-direction as one of three principal axes of the dielectric tensor $ \hat{\epsilon} $ (see below).
The amplitude $ {\boldsymbol{f}} ( x, y, z ) $ is written in terms of the
linear polarization basis: $ {\boldsymbol{f}} = {}^t ( f_1, f_2 ) = f_1 {\boldsymbol{e}}_1 + f_2 {\boldsymbol{e}}_2 $
with the mutual orthogonal basis $ {\boldsymbol{e}}_{ 1 , 2 } $.
Alternatively, we transform the basis vectors to the circular polarization basis:
$ {\boldsymbol{f}} = \psi_{1} {\boldsymbol{e}}_{+} + \psi_{2} {\boldsymbol{e}}_{-} $
with $ {\boldsymbol{e}}_{\pm} = \left( 1 / { \sqrt{2} } \right) ( {\boldsymbol{e}}_{1} \pm i {\boldsymbol{e}}_{2} ) $.
The corresponding two-component wave is given by
\begin{eqnarray}
  \psi
  \equiv
  \begin{pmatrix}
    \psi_{1} \\
    \psi_{2}
  \end{pmatrix}
  =
  \frac{ 1 }{ \sqrt{2} }
  \begin{pmatrix}
      1 & -i \\
      1 & i
  \end{pmatrix}
  \begin{pmatrix}
    f_{1} \\
    f_{2}
  \end{pmatrix}
  \equiv
  T
  \begin{pmatrix}
    f_{1} \\
    f_{2}
  \end{pmatrix}.
  \label{spinor}
\end{eqnarray}
Now we introduce the Stokes parameters, for which it is efficient to use the wave (\ref{spinor})
written in terms of the circular polarization basis \cite{Dennis,Kakigi} in contrast to the linear polarization basis,
e.g., Ref.~\cite{Brosseau}.
The former is particularly suitable to discuss the polarization singularity as is seen below.
Using the Pauli-spin $ \hat{\sigma}_{i} $, the Stokes parameters are written as
\begin{equation}
  S_i = \psi^{\dagger} \hat{\sigma}_{i} \psi,
  ~
  S_0 = \psi^{\dagger} {\boldsymbol{1}} \psi
  ~~~ {\rm{with}} ~~~
  ( i = 1 , 2, 3 )
\end{equation}
or explicitly,
\begin{equation}
  S_1 = \psi^{*}_{1} \psi_{2} + \psi^{*}_{2} \psi_{1} ,
  ~
  S_2 = i ( \psi^{*}_{2} \psi_{1} - \psi^{*}_{1} \psi_{2} ) ,
  ~
  S_3 = \psi^{*}_{1} \psi_{1} - \psi^{*}_{2} \psi_{2} .
\end{equation}
The vector $ {\boldsymbol{S}} ~( = ( S_1 , S_2 , S_3 ) ) $ satisfies the relation
$ S_{0}^{2} = S_{1}^{2} + S_{2}^{2} + S_{3}^{2} $, namely, $ S_{0} $
represents the field strength, $ S_{0} \equiv \left| {\boldsymbol{E}} \right|^{2} $.
Using the complex polar representation \cite{footnote}
\begin{equation}
  \psi_1 = \sqrt{ S_0 } \cos \frac{ \theta }{ 2 },
  ~~~
  \psi_2 = \sqrt{ S_0 } \sin \frac{ \theta }{ 2 } \exp \left[ i \phi \right],
  \label{spinor2}
\end{equation}
we obtain the polar form:
\begin{equation}
  {\boldsymbol{S}} = ( S_{0} \sin \theta \cos \phi ,~ S_0 \sin \theta \sin \phi ,~ S_0 \cos \theta )
  \label{stokes}.
\end{equation}
This parametrization is used for the pictorial representation of the polarization on the Poincar\'e sphere, which 
is given in Fig.1  \\

{\it Singular feature of the Stokes parameters}:
The expressions (\ref{spinor2}) and (\ref{stokes}) bear the polarization singularity in an apparent way.
This is shown by introducing the stereographic coordinate \cite{Arecchi}
\begin{equation}
  w = \frac{ S_1 + i S_2 }{ S_0 + S_3 } = \tan \frac{ \theta }{ 2 } \exp \left[ i \phi \right] ,
\end{equation}
which is the projection from the south pole.
From this expression one sees that the phase $ \phi $ cannot be defined at
the point $ w = 0 $ or $ \theta = 0 $, which corresponds to the Stokes vector $ S_3 = + S_0 $.
We have another form of the stereographic coordinate that means the projection from the north pole:
\begin{equation}
  w = \frac{ S_1 - i S_2 }{ S_0 - S_3 } = \cot \frac{ \theta }{ 2 } \exp \left[ - i \phi \right] ,
\end{equation}
for which the phase $ \phi $ is also indefinite at $ w = 0 $ or $ \theta = \pi $,
which corresponds to the Stokes vector $ S_3 = - S_0 $.

The above observation is reminiscent of the fact that the phase {\it loses} the meaning at the zero of wave function.
This implies that the two circular polarizations reveal the singularities of the polarized light, called the C-point.
On the other hand, at the equatorial circle $ \theta = \frac{\pi}{2} $, $ w = \exp[-i\phi] $
(or $ w = \exp[i\phi] $), there do  not reveal any singularities, and 
this circle divides the Poincar\'e sphere into two hemispheres corresponding to left-handed and right-handed  polarizations.
So we call the equatorial circle the L-line.

\begin{figure}[h]
  \begin{center}
    \includegraphics[width=45mm]{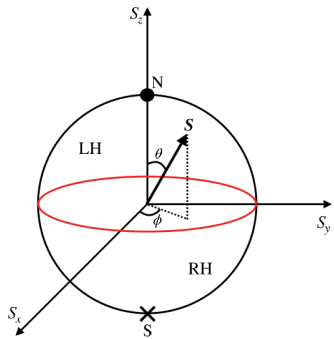}
   \caption{Schematic image of the Poincar\'e sphere.
    $ N $ and $ S $ denote left-handed and right-handed circular polarization, respectively.
   ``LH" and ``RH" mean the left-handed and ight-handed polarization, respectively.
    The red line (equator) is the linear polarization.
    }
    \label{fig:poincare}
  \end{center}
\end{figure}

\subsection{Evolution equation}

In the paraxial scheme, the Maxwell (Helmholz) equation is written for the electric field $ {\boldsymbol{E}} $:
\begin{eqnarray}
  \frac{ {\partial}^{2} {\boldsymbol{E}} }{ \partial z^{2} } + {\nabla}^{2} {\boldsymbol{E}}
  + \left( \frac{ \omega }{ c } \right)^{2} \hat{\epsilon} {\boldsymbol{E}} = 0 ,
  \label{one1}
\end{eqnarray}
where $ {\nabla}^{2} $ means the Laplacian in the transverse plane,
$ {\nabla}^{2} \equiv \frac{ {\partial}^{2} }{ \partial x^{2} } + \frac{ {\partial}^{2} }{ \partial y^{2} } $.
The dielectric tensor $ \hat{\epsilon} $ stands for the $ 2 \times 2 $ Hermitian matrix.

First we consider the special case that the Laplacian can be discarded, which
means that the modulation of light field $ {\boldsymbol{E}} $ in the transverse plane $ ( x , y ) $
changes slowly compared with the change in the longitudinal direction $ z $.
This procedure may be called the single mode approximation, for which the modulated amplitude
depends only on the propagation distance $ z $, namely, $ {\boldsymbol{E}} (z) $.
Hence we write
\begin{eqnarray}
  \frac{ {\partial}^{2} {\boldsymbol{E}} }{ \partial z^{2} }
  + \left( \frac{ \omega }{ c } \right)^{2} \hat{\epsilon} {\boldsymbol{E}} = 0 .
  \label{one2}
\end{eqnarray}
By applying the approximation of slowly varying amplitude (alias the envelope approximation)~\cite{Akhmanov},
namely, $ \left| \frac{ {\partial}^{2} {\boldsymbol{f}} }{ \partial z^{2} } \right| \ll k \left| \frac{ \partial {\boldsymbol{f}} }{ \partial z } \right| $,
we keep only the first derivative term
$ \frac{ \partial {\boldsymbol{f}} }{ \partial z } $.
Writing it in terms of the circular polarization basis,
we arrive at two-component nonlinear Schr\"odinger equation for $ \psi $:
\begin{eqnarray}
  i \lambda \frac{ \partial \psi }{ \partial z } = \hat{H} \psi .
  \label{wave1}
\end{eqnarray}
Here, $ \lambda \left( = \frac{ 2 \pi }{ k } \right) $ is the wave length.
The transformed Hamiltonian is given by
$ \hat{H} = - \frac{ \pi }{ n_{0} } T ( \hat{\epsilon} - n_{0}^{2} ) T^{-1} $.
In terms of the Pauli-spin we write
\begin{equation}
  \hat{H} = \alpha \hat{\sigma}_{1} + \beta \hat{\sigma}_{2} + \gamma \hat{\sigma}_{3} ,
\end{equation}
which may depend in general on the field $ (\psi^{\dagger}, \psi) $ as is shown below.
In what follows the wave equation (\ref{wave1}) is transcribed to the evolutional equation
for the Stokes parameters (see, e.g.,Refs. \cite{Sala,Tratnik,Kakigi}).
The concise way to carry this out is given by using the action principle \cite{Kra}.
That is, we introduce the action function $ I = \int L \, dz $ with the ``Lagrangian":
\begin{align}
  L
  & = \psi^{\dagger} \left( i \lambda \frac{ \partial }{ \partial z } - \hat{H} \right) \psi \nonumber \\
  & = - \frac{ S_0 \lambda }{ 2 } \left( 1 - \cos \theta \right) \dot{\phi} - H ( \theta, \phi ) .
\end{align}
The variational equation $ \delta I = 0 $ leads to
\begin{equation}
  \dot{\theta}
  =
  \frac{ 2 }{ S_{0} \lambda } \frac{ 1 }{ \sin \theta } \frac{ \partial H }{ \partial \phi },
  ~~
  \dot{\phi}
  =
  - \frac{ 2 }{ S_{0} \lambda } \frac{1}{\sin\theta}\frac{\partial H}{\partial\theta} .
\end{equation}
Using the Stokes vector, we have
\begin{equation}
  \frac{ d {\boldsymbol{S}} }{ dz }
  =
  - \frac{2}{ \lambda }
  \left( {\boldsymbol{S}} \times \frac{ \partial H }{ \partial {\boldsymbol{S}} } \right)
  \label{evolution}
\end{equation}
with the {\it expectation value} of the  Hamiltonian $ H = {\psi}^{\dagger} \hat{H} \psi $.
The evolution equation (\ref{evolution}) determines the orbit on the Poincar\'e sphere once
 the parameters $ ( \alpha, \beta, \gamma ) $ are provided. 

Following the conventional terminology,
the meaning for the parameters appearing in the Hamiltonian is explained as follows:
$ \alpha $  and $ \beta  $ represent the linear birefringence, whereas $ \gamma $ represents the optical activity or chirality.
The crucial point here is that these parameters depend on the spatial coordinate,
generally such that $ \alpha(x,y.z) $ etc.
In what follows, we  restrict the argument to the special case in which
these parameters have dependence only on the transversal coordinates, such that $ \alpha( x, y ) $.

\section{Trajectory of the polarization singularities}
\label{sec:trajectory}

Now we come to the first topic: to trace out the singularity of polarization in the linear birefringent medium.
The birefringence concerned here is not the one that is inherent in the nature of media,
e.g., the crystal structures, but it is caused by external electromagnetic field or self-induced birefringence,
which are discussed separately in what follows.

\subsection{Linear birefringence caused by an external Kerr effect}
\label{subsec:linearbiref}

This case is realized in such a way that the electric field is applied in the transverse plane:
$ {\boldsymbol{{\mathcal{E}}}} (x,y) = ( {\mathcal{E}}_{x}, {\mathcal{E}}_{y} ) $,
which is arranged to depend on the transversal coordinate $ (x,y) $.
In this way, the coordinates $ (x,y) $ could be treated as if
they were constant during manipulation of the evolution of the Stokes vectors.

As the dielectric tensor, we adopt the one coming from an external Kerr effect~\cite{LL},
namely,
\begin{equation}
  {\epsilon}_{ij} = n_{0}^{2} \delta_{ij} + G {\mathcal{E}}_{i} {\mathcal{E}}_{j} .
  \label{extKerr}
\end{equation}
Hence the wave equation for the amplitude $ {\boldsymbol{f}} $
(namely, the linear polarization basis) is written as
\begin{align}
  i \lambda \frac{ \partial }{ \partial z }
  \begin{pmatrix}
    f_1 \\
    f_2 \\
  \end{pmatrix}
  =
  \begin{pmatrix}
    a & b \\
    b & - a \\
\end{pmatrix}
\begin{pmatrix}
  f_1 \\
  f_2 \\
\end{pmatrix} .
\label{extKerr3}
\end{align}
with
$ a = \frac{ \pi }{ n_{0} } G \left[ \frac{ {\mathcal{E}}_{x}^{2} - {\mathcal{E}}_{y}^{2} }{ 2 } \right] $ and 
$ b = \frac{ \pi }{ n_{0} } G {\mathcal{E}}_{x} {\mathcal{E}}_{y} $.
We have here omitted the term arising from the conventional Kerr effect,
which is proportional to $ G \vert {\mathcal{E}} \vert^{2} $
with $ \vert {\mathcal{E}} \vert^2 = {\mathcal{E}}_{x}^{2} + {\mathcal{E}}_{y}^{2} $.
In terms of the circular polarization basis, this results in the Hamiltonian:
\begin{align}
  \hat{H}_l
  =
  a \hat{\sigma}_1 + b \hat{\sigma}_2
  \equiv
  \begin{pmatrix}
      0    &  a - ib \\
    a + ib &   0     \\
  \end{pmatrix} .
  \label{extKerr1}
\end{align}
Now it is instructive to examine the wave equation~(\ref{extKerr3}),
for which there are two independent solutions
$ f_{\pm} (z) = \exp \left[ \pm i \frac{ G }{2 n_{0} } \vert {\mathcal{E}} \vert^{2} k z \right] $.
By combining these with the reference plane wave, one gets the total polarization wave
\begin{equation}
  f_{\|} (z) = \exp \left[ i n_{\|} k z \right],
  ~~
  f_{\perp}(z) = \exp \left[ i n_{\perp} k z \right],
\end{equation}
which correspond to two refractive indices, namely:
$ n_{\|} = n_{0} + \frac{ G }{ 2 n_{0} } \vert {\mathcal{E}} \vert^2 $ and 
$ n_{\perp} = n_{0} - \frac{ G }{ 2 n_{0} } \vert {\mathcal{E}} \vert^2 $,
where the symbol $ \| $ means the polarization parallel to the applied field,
while $ \perp $ means the polarization perpendicular to the applied field~\cite{foot}.
As such the emergence of two refractive indices represents the literal meaning of birefringence,
which is contrast to the optical activity.

We now examine the evolution of the Stokes parameters.
The corresponding Hamiltonian reads $ H = a S_1 + b S_2 $, which is transformed to
\begin{equation}
  H = c S_{1}'
  \label{prime}
\end{equation}
by using the orthogonal transform:
\begin{align}
  S_{1}' = \cos \Theta S_1 + \sin \Theta S_2,
  ~~~
  S_{2}' = - \sin \Theta S_1 + \cos \Theta S_2,
  ~~~
  S_3' = S_3
  \label{ortho}
\end{align}
with $ \tan \Theta = \frac{b}{a} $.
Here, for the sake of simplicity of notation,
we write $ c = \sqrt{ a^2 + b^2 } $.
Then the evolution equation becomes
\begin{equation}
  \lambda \frac{ d S_{1}' }{ dz } = 0,
  ~~~
  \lambda \frac{ d S_{2}' }{ dz } = - 2 c S_{3},
  ~~~
  \lambda \frac{ d S_{3} }{ dz } = 2 c S_{2}'
\end{equation}
leading to
\begin{equation}
  S_{3} (z) = S_{0} \cos \left[ \frac{ 2c z }{ \lambda } \right],
  ~~~
  S_{2}' = - S_{0} \sin \left[ \frac{ 2c z }{ \lambda } \right]
\end{equation}
with the initial phase being chosen to be zero.

We summarize a type of polarization, ``quantization rule" and trajectory for each typical value of $ S_{3} (z) $
in Table~\ref{table:linear}.

\begin{table}[htb]
  \caption{The type of polarization, quantization rule and trajectory for each typical value of $ S_{3} (z) $ with integer $ n $.}
  \begin{tabular}{|c|c|c|c|} \hline
    The value of $ S_{3} (z) $ & Polarization type & Quantization rule & Trajectory \\ \hline \hline
    $ + S_{0} $ & Circular (Left-handed)  & $ \frac{ 2cz }{ \lambda } = 2 n \pi $ & C-point \\
    $ - S_{0} $ & Circular (Right-handed) & $ \frac{ 2cz }{ \lambda } = ( 2 n - 1 ) \pi $ & C-point \\ \hline
    $   0     $ & Linear & $ \frac{ 2cz }{ \lambda } = \left( n + \frac{1}{2} \right) \pi $ & L-line \\ \hline
  \end{tabular}
  \label{table:linear}
\end{table}

Thus, noting that
$ c ~( \equiv \sqrt{ a^2 + b^2 } ) \propto G ( {\mathcal{E}}_{x}^{2} + {\mathcal{E}}_{y}^{2} ) $,
the above rule forms a surface in $ ( x, y, z ) $ space.
From Table~\ref{table:linear}, for the particular case, $ n = 0 $, we
have $ {\mathcal{E}}_{x} = 0 $  and $ ~ {\mathcal{E}}_{y} = 0 $.
These two equations determine the point in the transversal plane $ ( x, y ) $.
In other words, the zero point of the external electric field forms a C-line in three dimensional space.
For nonzero $ n $, the corresponding trajectory forms the objects;
the C-points become the ``C-surface",
whereas the L-line the ``L-volume".
If cutting these objects by the plane $ z = const. $,
we have co-centric pattern according to the value $ n $.
These represent the L-surface and C-circle, so to speak,
which are arranged according to the numbering of $ n $.

\begin{figure}[h]
  \begin{center}
    \vspace{20mm}
    \hspace{-45mm}
    \includegraphics[width=\columnwidth]{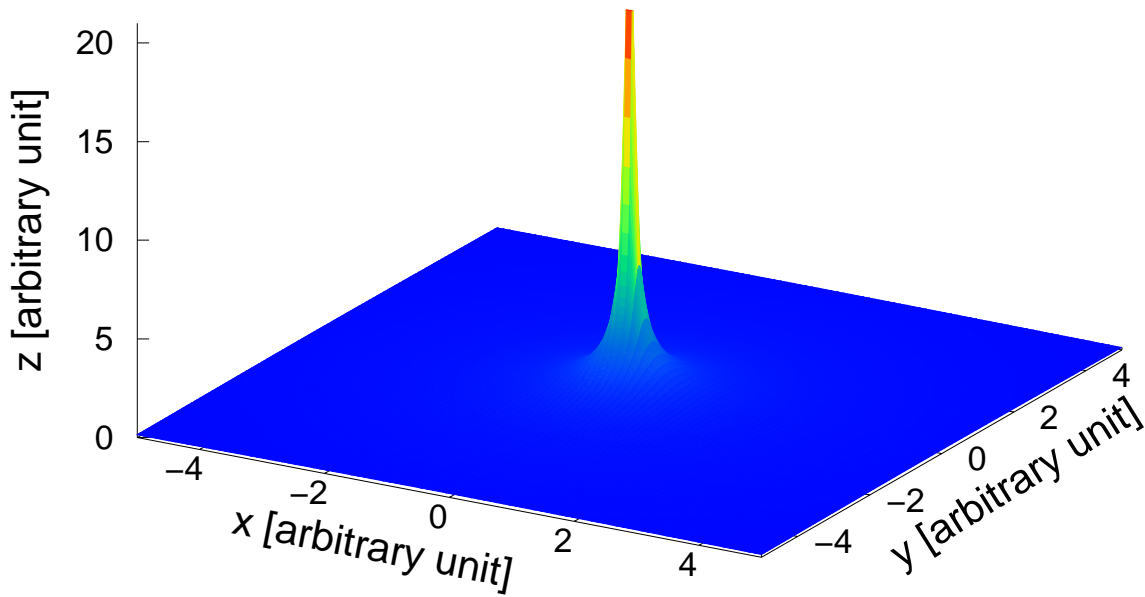}
    \vspace{-20mm}
    \caption{Outline of the polarization singularity: $ r z = const $.
    The $ z $ axis, namely $ r = 0 $, indicates the C-line.}
    \label{fig:outline_pol}
  \end{center}
\end{figure}

At this point it is natural to consider the case such that
$ \mathcal{E}  = \vert \mathcal{E}_{0} \vert  \sqrt{f(r)} \hat{r} $,
($ r = \sqrt{ x^2 + y^2 }, ~ \hat{r} = \frac{ {\boldsymbol{r}} }{ r } $),
namely, the radial vector in the transverse plane, which is axially symmetric with respect to the $ z $ axis.
In what follows we demonstrate the simplest choice, $ f(r) = r $.
The corresponding profile is given in Fig.~\ref{fig:outline_pol},
which represents the rotational body around $ z $-axis.
It is apparent that the $ z $ axis (that is $ r = 0 $) forms a C-line.

Furthermore the following point must be  mentioned:
The configurational structure of polarization singularity
can be controlled from an external condition through modulation of an electric field,
for example, by allowing the time variation of the amplitude $ {\mathcal{E}}_{0} (t) $.

{\it Remarks on the role of optical activity}:
In the above argument, it is essential that the Hamiltonian does not
include the term coming optical activity or chirality governed by $ S_3 $,
which causes $ S_3 $ to be invariant under Eq.~(\ref{ortho}).
So one may wonder how about the the effect of chirality or optical activity.
It is easy to examine this effect:
Instead of Eq.~(\ref{prime}) let us consider $ H = a S_1 + \gamma S_3 = \Gamma S_{1}' $
with $ \gamma = v {\mathcal{B}} $ ($ {\mathcal{B}} $ is the magnetic field and $ v $ is the Verde constant),
$ \Gamma = \sqrt{ a^2 + {\gamma}^2 } $.
One has the orthogonal transform corresponding to Eq.~(\ref{ortho})
\begin{align}
  S_{1}' = \cos \Phi S_{1} + \sin \Phi S_{3},
  ~~~
  S_{3}' = - \sin \Phi S_{1} + \cos \Phi S_{3},
  ~~~
  S_{2}' = S_{2}
  \label{ortho2}
\end{align}
with $ \tan \Phi = \frac{\gamma}{a} $.
Then the evolution equation becomes
$ \lambda \frac{ d S_{1}' }{ dz } = 0 ,~ \lambda \frac{ d S_{2} }{ dz } = - 2 \Gamma S_3' ,~ \lambda \frac{ d S_{3}' }{ dz } = 2 \Gamma S_2 $,
from which one gets
$ S_3(z) = C \sin \Phi + S_0 \cos \Phi \cos \left[ \frac{2}{\lambda} \Gamma z \right]  $,
where $ C $ is the constant of motion $ S_1' = C $.
The trajectory of the C-line (or L-surface) is thus derived from
\begin{align}
  \cos \left[ \frac{2}{\lambda} \Gamma z \right]
  & =
  \pm \frac{1}{ \cos \Phi } - \frac{ C \tan \Phi }{ S_0 } ~~({\rm{Circular}}), \nonumber \\
  & =
  - \frac{ C \tan \Phi}{ S_0 } ~~({\rm{Linear}}) .
\end{align}
From this expression one sees that the additional term,
which depends on the angle $ \Phi $ as well as arbitrarily chosen constant of motion $ C $,
{\it violates} a simple quantization rule leading to the polarization singularities.
This fact suggests that the optical activity plays a role of {\it obstacles} for forming the polarization singularities.
The exceptional case is that; $ \Phi = 0 $ (that means $ \gamma $ vanishes),
for which the problem is reduced to the case of the pure birefringence.

As an extreme case, we consider the pure optical activity, for which the evolution equation reads
$ \dot{S}_{1} = - \frac{2}{\lambda} \gamma S_2 ,~ \dot{S}_{2} = \frac{2}{\lambda} \gamma S_1 ,~ \dot{S}_{3} = 0 $.
From this we have the constant of motion $ S_3 = C $.
Here the points $ C = \pm S_0 ,~ 0 $; the circular and linear polarization should be careful.
Namely, the circular polarization consists only of ``isolated points" on a Poincar\'e
sphere arbitrarily chosen in 3-dimensional space,
whereas the linear polarization describes the equatorial circle on the same sphere.

\subsection{Including the nonlinear birefringence}

Turning to the nonlinear Kerr media, it is known that the dielectric tensor is modified
to be expressed in terms of the complex components of the spinor~\cite{LL,Maker}.
Namely this is given as
\begin{eqnarray}
  \epsilon_{ij} = n_{0}^{2} \delta_{ij} + g \left( E_{i}^{*} E_{j} + E_{j}^{*} E_{i} \right) .
\end{eqnarray}
The corresponding Hamiltonian operator becomes
\begin{equation}
  \hat{V}_{nl}
  =
  - g
  \begin{pmatrix}
    - \frac{ \vert \psi_1 \vert^2 - \vert \psi_2 \vert^2 }{ 2 } & \psi_{2}^{*} \psi_{1} \\
    \psi_{1}^{*} \psi_{2} & \frac{ \vert \psi_1 \vert^2 - \vert \psi_2 \vert^2 }{ 2 }
  \end{pmatrix} ,
  \label{nlKerr}
\end{equation}
which results in the expectation value:
$ \psi^{\dagger} \hat{V}_{nl} \psi = - \frac{g}{2} ( S_1^2 + S_2^2 - S_3^2 ) \equiv g S_3^2 $ up to additional constant.
By taking account of this term, the Hamiltonian (\ref{prime}) is modified as
\begin{equation}
  H = c S_{1}' + g S_{3}^2 .
\end{equation}
By this modified Hamiltonian, the equation of motion turns out to be~\cite{Sala,Kra}
\begin{equation}
  \lambda \frac{ d {\boldsymbol{S}}' }{dz}
  =
  \begin{pmatrix}
    - 4 g S_{2}' S_{3} \\
    - 2 c S_{3} + 4 g S_{1}' S_{3} \\
    2 c S_{2}'
  \end{pmatrix},
\end{equation}
for which we have the two constants  of motion:
\begin{equation}
  S_{1}'^{2} + S_{2}'^{2} + S_{3}^{2} = S_{0}^{2},
  ~~~
  c S_{1}' + g S_{3}^{2} = E .
\end{equation}
One sees that by  choosing  the energy constant as $ E = g S_{0}^{2} $,
the equation of motion is derived for the third component $ S_3 $.
According to the ratio between the linear birefringence and nonlinear coupling, namely,
$ \kappa = \frac{ g S_{0}^{2} }{ c } $ (called {\it moduli parameter}), we have two cases of equations for $ S_3 $;
(A) $ c > g S_{0}^{2} $ and (B) $ c < g S_{0}^{2} $~\cite{footnote1}.
By putting $ S_3 = S_{0} X $ and $ \tau = \frac{ 2c z }{ \lambda } $,
these are written as
\begin{align}
  \left( \frac{ dX }{ d \tau } \right)^{2}
  & =
  \left( 1 - X^{2} \right) \left( 1 - {\kappa}^{2} + {\kappa}^{2} X^{2} \right)
  ~~{\rm{for ~case ~(A) }},  \\
  \left( \frac{ dX }{ d \tau } \right)^{2}
  & =
  {\kappa}^{2} \left( 1 - X^{2} \right) \left( X^{2} - \bar{\kappa}^{2} \right)
  ~~{\rm{for ~case ~(B) }}
\end{align}
with $ \bar{\kappa} = \sqrt{ 1 - \frac{1}{ {\kappa}^{2} } } $.
Hence the corresponding solution of $ S_3 $ are given by
\begin{align}
  S_3 (\tau) & = S_0 \, {\rm{cn}} \left( \tau , \kappa \right)
  ~~{\rm{for ~case ~(A) }}, \\
  S_3 (\tau) & = S_0 \, {\rm{dn}} \left( \kappa \tau , \bar{\kappa} \right)
  ~~{\rm{for ~case ~(B) }},
\end{align}
where $ {\rm{cn}} $ and $ {\rm{dn}} $ represent the Jacobi elliptic functions.

The former solution indicates $ S_3 $ oscillates between $ - S_0 $ and $ S_0 $, while
the latter exhibits the small oscillation in the vicinity of the north (or south) poles;
this feature is implied from $ \bar{\kappa} S_0 \leq \vert S_3 \vert \leq S_0 $.
In what follows we give analysis for the trajectory of the C-points and L-lines,
$ S_3 (\tau) = \pm S_0 ,~ 0  $.

For case (A), $ \kappa < 1 $, we summarize a type of polarization, ``quantization rule" and trajectory for each typical value of $ S_{3} (z) $
in Table~\ref{table:nonlinear}.
\begin{table}[htb]
  \caption{The type of polarization, quantization rule and trajectory for each typical value of $ S_{3} (z) $ with integer $ n $.}
  \begin{tabular}{|c|c|c|c|} \hline
    The value of $ S_{3} (z) $ & Polarization type & Quantization rule & Trajectory \\ \hline \hline
    $ + S_{0} $ & Circular (Left-handed)  & $ \frac{ 2cz }{ \lambda } = 4 n K(\kappa) $ & C-point \\
    $ - S_{0} $ & Circular (Right-handed) & $ \frac{ 2cz }{ \lambda } = 2 n K(\kappa) ~~( \neq 4nK ) $ & C-point \\ \hline
    $   0     $ & Linear & $ \frac{ 2cz }{ \lambda } = ( 2n - 1 ) K(\kappa) $ & L-line \\ \hline
  \end{tabular}
  \label{table:nonlinear}
\end{table}

In Table~\ref{table:nonlinear}, $ K(\kappa) $ denotes the period of the $ {\rm{cn}} $ function,
which is given by the complete elliptic integral:
\begin{equation}
  K (\kappa)
  =
  \int_{0}^{1} \frac{ d \tau }{ \sqrt{ ( 1 - {\tau}^{2} )( 1 - {\kappa}^{2} {\tau}^{2} ) } } .
\end{equation}
The contents of the Table~\ref{table:nonlinear} correspond to 
the contents  obtained for the case of pure linear birefringence.

\begin{figure}[h]
  \begin{center}
    \vspace{20mm}
    \hspace{-45mm}
    \includegraphics[width=\columnwidth]{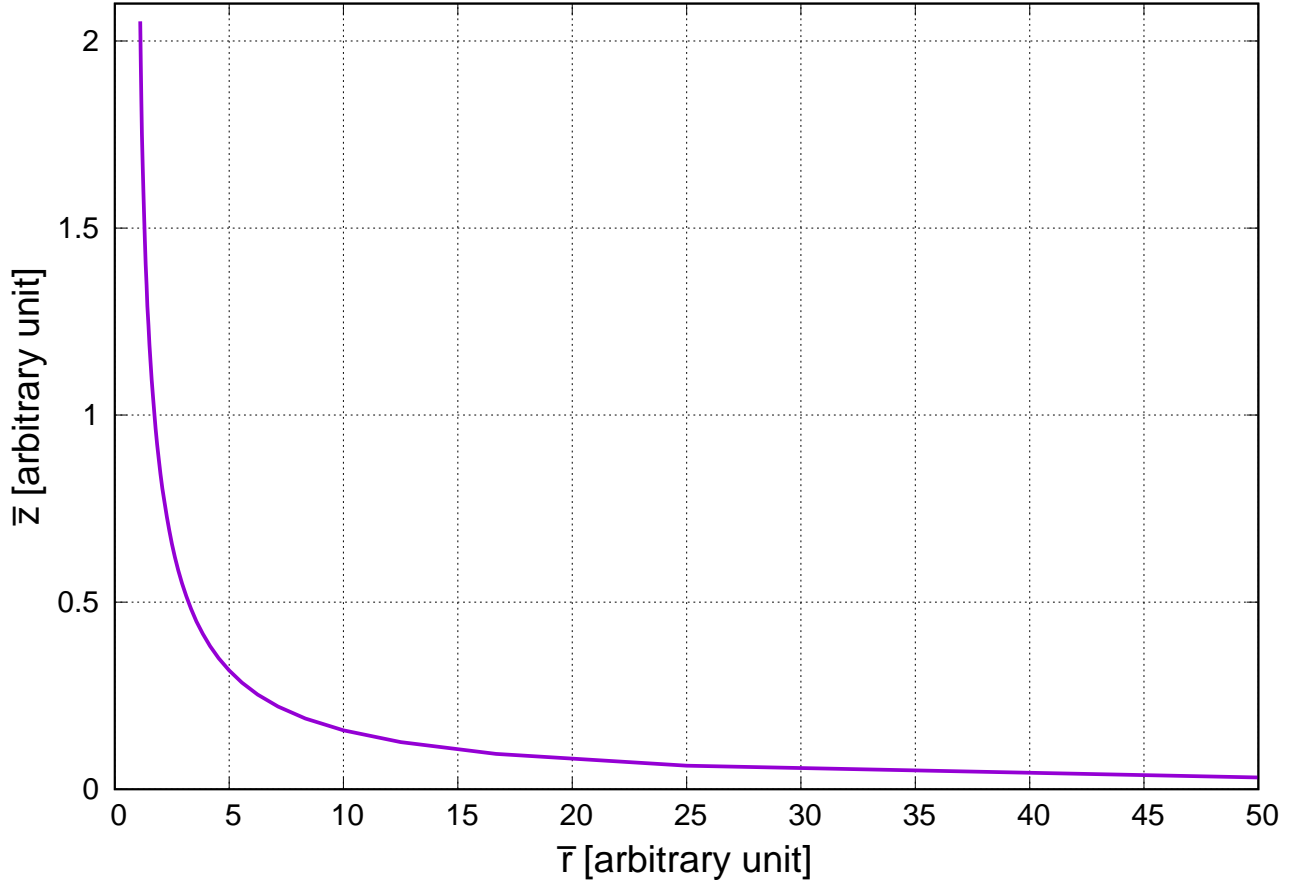}
    \vspace{-20mm}
   \caption{The trajectory in the $ ( \bar{r} , \bar{z} ) $ plane.
   }
    \label{fig:trajectory}
  \end{center}
\end{figure}

The trajectory revealing these singularities is rather complicated to display,
because of the highly sophisticated nature of the elliptic functions.
Here we  note that $ n = 0 $ is not allowed, because it yields
$ c \propto G \left( {\mathcal{E}}_{x}^{2} + {\mathcal{E}}_{y}^{2} \right) = 0 $,
but this contradicts to the constraint $ c > g S_0^2 > 0 $.
In other words, there does not appear the case that
$ {\boldsymbol{\mathcal{E}}} = ( {\mathcal{E}}_{x} , {\mathcal{E}}_{y} ) $
vanishes leading to the C-line.
This feature is in contrast to the case of the linear birefringence.

We  discuss the trajectory
of left-handed circular polarization
by adopting the same profile as
the linear birefringence for the applied external field, namely
$ \vert {\mathcal{E}} \vert^{2} = \vert {\mathcal{E}}_{0} \vert^{2} r $.
We take up the case $ n = 1 $, that is, the period $ 4 K(\kappa) $.
Then, we write
\begin{eqnarray}
  \bar{r}
  \left( \equiv \frac{ \pi }{ 2 n_{0} } \frac{ G \vert {\mathcal{E}}_{0} \vert^{2} r }{ g S_{0}^{2} } \right)
  =
  \frac{1}{\kappa} .
  \label{r}
\end{eqnarray}
On the other hand, the trajectory
of left-handed circular polarization \color{black}
is also written as
\begin{eqnarray}
  \bar{z}
  \left( \equiv \frac{ 2 g S_{0}^{2} }{ {\lambda} } z \right)
  =
  4 \kappa K( \kappa ) .
  \label{z}
\end{eqnarray}
With the aid of these two relations, one can eliminate the moduli parameter $ \kappa $ to
get a trajectory in the $ ( \bar{r} , \bar{z} ) $ plane (see Fig.~\ref{fig:trajectory}).
Note that in the limit $ \kappa \rightarrow 0 $ ($ g \rightarrow 0 $) we recover the trajectory for the linear birefringence.

Now turning to the case (B): $ \kappa > 1 $, we see that there appear two orbits, for which we give a brief sketch.
These are determined by $ {\rm{dn}} = \pm 1 $, which turn out to be
$ \frac{ 2cz }{ \lambda } = 4 K( 1 / \kappa ) $ for $ S_3 = + S_0 $ and
$ \frac{ 2cz }{ \lambda } = 2 K ( 1 / \kappa ) $ for $ S_3 = - S_0 $.
There are no orbits for the linear polarization, $ S_3 = 0 $,
because of the $ {\rm{dn}} $ function takes never zero.

{\it Remarks}:
If the term coming from the optical activity is added to the Hamiltonian,
the equation of motion is not reduced to a simple form that is described by the third component of the Stokes parameters.
This implies that one has the same problem as in the linear birefringence case;
that is, the optical activity destroys a simple quantization rule
to extract the configurational structure revealing the polarization singularities.

\section{Non-linear field and Optical skyrmion}
\label{sec:skyrmion}

We now consider the case that the polarized field extends over the transversal plane;
the field is described by $ \psi ( x, y, z ) $, that is,
the Laplacian is recovered~\cite{Kra2}.
\begin{eqnarray}
  i \lambda \frac{ \partial }{ \partial z }
  \begin{pmatrix}
    \psi_1  \\
    \psi_2
  \end{pmatrix}
  =
  \left( - \frac{ {\lambda}^{2} }{ 2 n_{0} } {\nabla}^{2} {\boldsymbol{1}} + \hat{V}_{nl} \right)
  \begin{pmatrix}
    \psi_1  \\
    \psi_2
  \end{pmatrix}
  \label{tcns}
\end{eqnarray}
with $ {\boldsymbol{1}} $ being the $ 2 \times 2 $ unit matrix.
The corresponding Hamiltonian is written in terms of
the angular field $ ( \theta(x,y) , \phi(x,y) ) $~\cite{Kra2}:
\begin{equation}
  \tilde{H}
  =
  \int \left[ \frac{ S_{0} {\lambda}^{2} }{ 2 n_{0} }
    \left\{ \frac{1}{4} \left( {\boldsymbol{\nabla}} \theta \right)^{2}
      + \sin^{2} \frac{ \theta }{ 2 } \left( {\boldsymbol{\nabla}} \phi \right)^{2} \right\}
      + g S_{0}^{2} \cos^{2} \theta \right] \, d^{2} x ,
  \label{field}
\end{equation}
where we have omitted an additional constant term.
Then the problem is what is expected for the polarization singularity in the framework of this field equation.
A plausible way to answer this question may be given by utilizing the special solution that is inherent in the 
nonlinear field equation (\ref{tcns}).
If we consider the characteristics of the nonlinear birefringence,
this can be achieved by the vortex type solution,
which is called the optical skyrmion in terms of the current terminology~\cite{Moon}.
By this specific solution the structure of the polarization singularity of
the C-line and the L-surface could be naturally incorporated.
Actually the same solution has been discussed in the previous work~\cite{Tsuchida},
but the argument from the polarization singularity remains untouched. 
Here we  give a discussion from this renewal point of view.

To construct the special solution we need to settle its profile.  First,
taking into account the topological characteristics,
the phase function is chosen such that $ \phi = n \tan^{-1} \left( \frac{y}{x} \right) $,
with $ n = 1, 2, \cdots $, being the winding number, which follows the idea of the usual vortex~\cite{footnote2}.
Next the angular function $ \theta $, which just determines the profile of the skyrmion,
is chosen as a function of the radial variable $ r ~( = \sqrt{ x^2 + y^2 } ) $.
By keeping these characteristics in mind, we write 
the Hamiltonian (\ref{field}) in terms of the field $ \theta(r) $:
\begin{eqnarray}
  H
  =
  \frac{ S_{0} {\lambda}^{2} }{ 8 n_{0} }
  \int \left[ \left\{ \left( \frac{ d \theta }{ dr } \right)^{2}
          + \frac{ 4 n^{2} }{ r^{2} } \sin^{2} \frac{ \theta }{ 2 } \right\} + g' \cos^2 \theta \right] r \, dr .
\end{eqnarray}
Here, we put $ g' = { 8 g n_{0} S_{0} \over {\lambda}^{2} } $, which means that
$ g' $ takes a positive value.
The profile function $ \theta(r) $ may be derived from the extremum of $ H $,
namely, the Euler-Lagrange equation leads to
\begin{eqnarray}
  \frac{ d^{2} \theta }{ d {\xi}^{2} } + \frac{ 1 }{ \xi } \frac{ d \theta }{ d \xi }
    - \frac{ n^{2} }{ {\xi}^{2} } \sin \theta + \frac{ 1 }{ 2 } \sin 2 \theta = 0 ,
\end{eqnarray}
where we adopt the scaling of the variable: $ \xi = \sqrt{g'} r $.
An efficient way to extract the polarization singularity
is given by checking the behavior of $ \theta(\xi) $
at specific boundary conditions at $ \xi = 0 $ and $\xi = \infty$.

The behavior near the origin $ \xi = 0 $ is controlled by the differential equation that
serves as the Bessel equation;
thus we obtain $ \theta(\xi) \simeq J_{n} (\xi) $, which satisfies $ \theta(0) = 0 $.
This means that the optical state is left-handed circular polarization at the origin.
On the other hand, the behavior at $ \xi = \infty $ is seen
by checking the stationary feature of the solution;
if putting $ \theta(\xi) = \frac{ \pi }{2} + \alpha $, with $ \alpha $ being the infinitesimal deviation,
then we have the linearized equation  $ {\alpha}'' - \alpha \simeq 0 $
near $ \xi = \infty $, which results in $ \alpha \simeq \exp [ - \xi ] $.
This means that the solution should approach to $ \theta(\infty) = \frac{ \pi }{2} $.
Thus, the state becomes linear polarization at infinity.

\begin{figure}[h]
  \begin{center}
    \includegraphics[width=80mm]{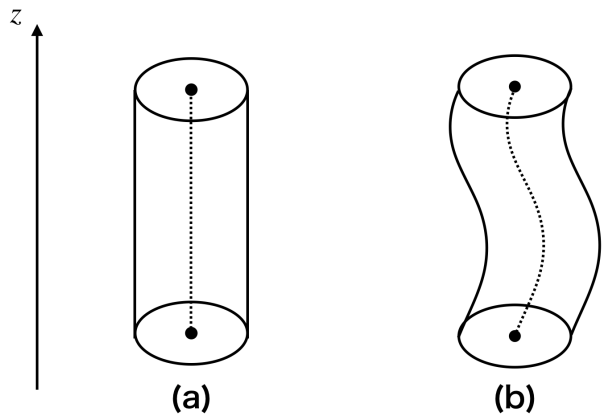}
    \caption{Conceptual image of the skyrmion tube.
    The surface of the tube, namely the L-surface, corresponds to the equator of the Poinc{\`a}re sphere.
    The center of the tube represents the C-line.
    (a) No external force, (b) Including external force.
    }
    \label{fig:PS}
  \end{center}
\end{figure}

From the above construction of the optical skyrmion,
we see that this reveals a three dimensional object,
if it is extended in the propagation direction.
The center line forms a C-line, whereas the boundary forms a L-surface.
From the pictorial viewpoint (see Fig.~\ref{fig:PS}),
there is somehow similarity with Fig.~\ref{fig:outline_pol}
which profiles the polarization singularity for the linear birefringence.
Specifically, to be noted is that the C-line, the central axis $ r = 0 $,
just corresponds to the center line of the optical skyrmion.
The C-line may be deformed by an action coming from externally driving (pinning) force.
In the presence of such external force,
the behavior of C-line becomes  like that in Fig.~\ref{fig:PS} (b),
which is rather intricate in general to extract the analytical solution for the behavior of the C-line.

Apart from the configurational aspect of the optical skyrmion, which is quoted as
$ \psi_{\rm{SK}} $,
the more interesting aspect is its dynamical content,
following the original idea of Skyrme~\cite{Skyrme}.
Namely, the tube domain carries a {\it kinetic energy} along with the propagation direction $ z $.
This can be added to Eq.~(\ref{field}) as a form that is written in terms of the spinor wave $ \psi $:
$ T \propto \left| \frac{ \partial \psi }{ \partial z } \right|^{2} $
which is the energy corresponding to the second derivative
$ \frac{ {\partial}^{2} {\boldsymbol{E}} }{ \partial z^{2} } $ of the starting Maxwell equation (\ref{one1}),
that is discarded in the envelope approximation.
(Note that in terms of the Stokes parameter, it is written as
$ T \propto \left( \frac{ {\partial} {\boldsymbol{S}} }{ \partial z } \right)^{2} $).
This term implies that the Stokes vector gives rise to the structure of
the {\it spinning} (or rotational) degree together with the translational mode.
Indeed the quadratic term of $ \psi $ may bring about the kinetic energy, that is,
if we write the wave incorporating the {\it collective degree}
$ \Phi $ and ${\boldsymbol{R}} $
(here, $ \Phi $ means the global phase,
and $ {\boldsymbol{R}} $ is the coordinate, which represents the trajectory of the C-point):
\begin{eqnarray}
  \psi (x,y,z)
  =
  \exp \left[ i \Phi(z) \right]
  \psi_{\rm{SK}} \left[ {\boldsymbol{x}} - {\boldsymbol{R}}(z) \right] ,
\end{eqnarray}
it follows that $ T \propto A \dot{\Phi}^{2} + B \dot{\boldsymbol{R}}^{2} $;
the sum of the spinning and translational energy of the optical skyrmion.
Here we note that there appears an essentially different term inherent in the spin degree;
called the {\it gyration term} that is proportional to $ \dot{\boldsymbol{R}} $,
which comes from the first order derivative of the Stokes parameter~\cite{Tsuchida}.

Here a remark is given on the stability of the skyrmion against the various kind of external perturbations.
In order to examine this feature, let us consider the dynamical and topological aspect.
The dynamical stability of the skyrmion will be carried out by the eigen-mode analysis
for the linearized equation, which describes the deviation from the solution  $ (\theta, \phi) $.
It is crucial that, in the present context, without detailed analysis of this linearized equation,
the feature of polarization singularity is kept even in the presence of external perturbation.
That is,
there is a stability from a topological aspect, which may be described by the topological charge.
The main feature of the skyrmion can be topologically  protected against the external disturbance~\cite{Skyrme}.
The details of the eigen-mode analysis will be left for future study.

We further mention the other aspect of skyrmion: e.g., the occurrence of multiple skyrimions
in analogy with the multiple vortices that appear in the conventional single component
nonlinear Schr{\"o}dinger equation.
Such aspect could not be expected in the dynamics of the single mode Stokes parameters.

\section{Discussion and summary}

{\it  A brief remark on a possible experimental realization}:
(i)  For the single mode case, 
the first thing  is to arrange the linear birefringence by using the electric field.
This is provided in such a way that the electric charge is uniformly
distributed on a ``line" (or rod) so as to be in an axially symmetric way about the line,
then an axially symmetric electric field may be produced according to the law of electro-statics,
which is as given in Sec.~\ref{subsec:linearbiref}.
This can be extended to the case where there is nonlinear birefringent media;
that is, the charged line is inserted in the media.
Having given the setting, one can observe the C-line (L-line) {\it in principle}. Namely, appropriately arranged
polarized light is transmitted through the birefringent apparatus, then the output polarization is expected to be
observed with the aid of an analyzer (e.g. Berry and Dennis \cite{Berry}).
Here particularly mentioned is the time variation of the birefringence,
which is controlled by the time varying amplitude of the electric field.
According to the formula given in Table~\ref{table:linear} and \ref{table:nonlinear},
the period $ c $, which stands for the change of the C-point, depends on time explicitly.
The time variation would give rise to a  direct modulation of the C-point for fixed coordinate $ ( r, z) $.
A similar argument can be applied to the case for the L-lines.

(ii) On the other hand, for the case of the optical skyrmion, the following setting will be supposed.
According to the experimental setting of Ref.~\cite{Swartz},
let us consider the circular polarized beam passing through a nonlinear birefringent media.
During the propagation through the media, the vortex (skymion) will be produced.
The crucial point is that the vortex mode somehow forms the {\it wave guide (tube)}.
If probe light that is  circularly  polarized  is injected through this
tube, then one could observe the output circularly polarized light by the analyzer at the end.
So the actual occurrence of this event may be  evidence of the optical skyrmion.  \\

{\it Summary}. 
The circular and linear polarization play a peculiar role in the light polarization, which
is represented symbolically by the terminology of ``polarization singularity."
In particular the circular polarization is characterized by the feature that
the azimuthal angle $ \phi $ is indefinite at the north and south poles.
Under this standpoint we have explored the polarization singularity from two aspects:
On the one hand, the configurational structure of polarization singularity, which is manifested by the
single mode Stokes vector, is given by the trajectory satisfying $ S_3(z) = \pm S_0 $ and $ ~0 $ that
is induced by the linear as well as nonlinear birefringence.
Here particularly mentioned is that the optical activity plays a role in preventing 
the formation of polarization singularity.
The polarization singularity may be manipulated by modulating the external electric field.
On the other hand, as a supplement to the single Stokes parameters,
we have discussed the nonlinear field by taking account of the transverse coordinate,
which generates a spontaneous structure of the polarization singularities in the form of optical skyrmions.
This provides a self-organized polarization singularity that is caused by the collaborated
effect of nonlinear birefringence and the field effect extended over the transverse plane.
A further analysis would be expected to reveal hidden aspects of polarization optics
on the basis of the attempt presented here.

\acknowledgments

The work of S. Tsuchida is supported in part by Grant-in-Aid for JSPS Research Fellows, No. 20J00978.


\begin{thebibliography}{13}
\bibitem{footnote}
  The angular representation adopted here is different from that of \cite{Dennis}.
  We use the standard spinor parametrization \cite{LL2}, which leads to the
  natural definition of the spin vector (Stokes parameters) given below.
  See also \cite{Arecchi}.
\bibitem{foot}
  An alternative expression for $ n_{\|} $ snd $ n_{\perp} $ are
  $ n_{\|} = n_{0} ,~ n_{\perp} = n_{0} - \frac{G}{ n_{0} } \vert {\mathcal{E}} \vert^{2} $
  which are simply obtained by adding the contribution from the isotropic term
  $ \frac{ G }{ 2 n_{0} } \vert {\mathcal{E}} \vert^{2} $.
  See Ref.~\cite{LL}.
\bibitem{footnote1}
  The third case $ c = g S_0^2 $ gives $ S_3 = {\rm{sech}} \tau $,
  which means a localized soliton mode.
  This case is apparently irrelevant.
\bibitem{footnote2}
  It is known that the conventional skyrmion is characterized by the topological invariant (charge).
  For the present optical skyrmion we can construct the correspondent quantity,
  which is given by   $ \sigma =  \int ( {\boldsymbol{\nabla}} \times {\boldsymbol{v}} ) \, d^{2} x $
  where $ {\boldsymbol{v}} = ( 1 - \cos \theta ) {\boldsymbol{\nabla}} \phi $ is the velocity field.
  It turns out to be  $ \int \sin \theta \, d \theta \wedge d \phi $, which means
  the mapping from the compactified plane $ \{ {\boldsymbol{R}} \cup {\infty} \} $ to the Poincar{\'e} sphere.
  See, e.g., Refs. \cite{Ho,Kura}.
\end{thebibliography}
\end{document}